\documentclass[12pt,preprint]{aastex}  
\def\alwaysmath#1{\ifmmode{#1}\else{$#1$}\fi}

\def\ltsima{$\; \buildrel < \over \sim \;$}  
\def\gtsima{$\; \buildrel > \over \sim \;$}  
\def\lsim{\lower.5ex\hbox{\ltsima}}  
\def\gsim{\lower.5ex\hbox{\gtsima}}  
  
\slugcomment{In press in  ApJ Letters}  
  
\begin{document}

\title{
Discovery of an anomalous Sub Giant Branch in the  
Color Magnitude Diagram of $\omega$ Centauri    
\footnote{Based on  
observations collected at the Very Large Telescope  
of the European Southern Observatory, Cerro Paranal, Chile,   
within the observing programme 68.D-0332.  
Also based on observations with the NASA/ESA HST, obtained at  
the Space Telescope Science Institute, which is operated by AURA, Inc.,  
under NASA contract NAS5-26555.  
}}  
  
\author{Francesco R. Ferraro\altaffilmark{}  
\footnote{Dipartimento di Astronomia Universit\`a   
di Bologna, via Ranzani 1, I--40127 Bologna, Italy,  
ferraro@bo.astro.it},   
Antonio Sollima\altaffilmark{2,}\footnote{INAF-Osservatorio   
Astronomico di  Bologna,  
 via Ranzani 1, I--40127 Bologna,  Italy,
 antonio@omega.bo.astro.it, pancino@bo.astro.it, 
 bellazzini@bo.astro.it,origlia@bo.astro.it},  
Elena Pancino\altaffilmark{3}, Michele Bellazzini\altaffilmark{3},  
Oscar Straniero\footnote{INAF-Osservatorio Astronomico di Teramo, 
I-64100 Teramo, Italy, straniero@te.astro.it},  
Livia Origlia\altaffilmark{3},  
Adrienne M. Cool\footnote{Department of Physics and Astronomy, 
San Francisco State University, 1600 Holloway
Avenue, San Francisco, CA 94132, USA,  cool@sfsu.edu }  
}

\begin{abstract}  
  
Using deep high-resolution multi-band images taken with the   
{\it Very Large Telescope} and the Hubble Space Telescope, we  
discovered a new anomalous sequence in the Color Magnitude  
Diagram of $\omega$~Cen. This feature appears as a narrow,  
well-defined Sub Giant Branch (SGB-a), which merges into the Main  
Sequence of the dominant cluster population at a  magnitude 
significantly fainter than the cluster Turn-Off (TO).  
  
The simplest hypothesis assumes that the new feature is the
extension of the anomalous Red Giant Branch (RGB-a) 
metal-rich population discovered by
Lee et al. (1999) and Pancino et al. (2000). 
However, under this assumption 
the interpretation of the SGB-a 
does not easily fit into the context of a 
self-enrichment scenario within $\omega$~Cen. In fact, 
its TO magnitude, shape and extension 
are not compatible with a young, metal-rich 
population, as required by the self-enrichment process.
The TO level of the SGB-a suggests indeed an age as old as
the main cluster population, further supporting  
the extra-cluster origin of the most metal rich stars, 
as suggested by Ferraro, Bellazzini \& Pancino (2002). 
Only accurate measurements of radial velocities and  
metal abundances for a representative sample of stars  
will firmly establish whether or not the SGB-a is 
actually related to the RGB-a and will finally shed light on the
origin of the metal rich population of $\omega$~Cen.
 
\end{abstract}  
   
   
\section{Introduction}   
\label{sec:intro}  
  
The observational facts collected over the last 40 years have  
demonstrated that the most massive ($M\sim  
5\times10^6~M_{\odot}$) star cluster in the Halo of the Galaxy,  
$\omega$~Centauri, also shows the most surprising properties  
ever observed in a globular cluster. It is the only known  
Galactic globular with a clear metallicity spread   
(Norris \& Da Costa, 1995; Smith  
et al., 2000). Even more interesting, the abundance variations  
appear to be correlated with some peculiar structural and  
kinematical properties (Norris et al. 1997; Pancino et al.  
2003).   
  
Recent wide field photometric studies (Lee et al. 1999;
Pancino et al. 2000) have identified  a
previously unknown anomalous RGB sequence (RGB-a), clearly
separated from the bulk of the RGB population. The RGB-a photometric
properties suggest a very high metal content, confirmed by
spectroscopic follow-up ([Fe/H]$\simeq-0.6$; Pancino et
al. 2002; Origlia et al. 2003). 

In a self-enrichment
scenario, the RGB-a would belong to the last burst of star
formation and should thus be younger than other populations
in $\omega$~Cen. However, using the photometry by Pancino
et al. (2000) and the proper motions of van Leeuwen et al
(2000), Ferraro, Bellazzini \& Pancino (2002) showed that the
RGB-a stars have a coherent bulk motion with respect to the
other cluster stars, suggesting that the RGB-a could be a
captured component, now trapped in the potential well of
the cluster. This evidence could eventually fit into
a different scenario  
where the cluster is the remnant
(i.e., the nucleus) of a disrupting 
dwarf galaxy (see Bekki \& Freeman, 2003), not
dissimilar from the Sagittarius case (Majewski et al.
2000). 
 
In this letter, we present the discovery of an additional  
anomalous sequence in the Color Magnitude Diagram (CMD) of $\omega$~Cen: 
a narrow Sub Giant Branch (SGB-a)  
which merges into the Main Sequence (MS) of the cluster with a  
Turn-Off (TO) significantly fainter  than the
dominant cluster population. This new feature adds  
another piece to the complex $\omega$~Cen puzzle.  
  
\section{Observations and data analysis}  
\label{sec:obs}  
   
The photometric data consist of a set of high  
resolution images obtained with FORS1 at the VLT.  
All images were obtained (in service mode) between April and  
October 2002, under very good seeing conditions  
(FWHM$=0\farcs4-0\farcs9$), using the high-resolution FORS1  
imaging mode ($0.1\farcs/$pix) that gives a field of view of  
$3\farcm4 \times 3\farcm4$. We covered a global $\sim 9\farcm  
\times 9\farcm$ area with a mosaic of $3\times3$ pointings   
around the cluster center. Each pointing
consists  of four deep B, V and I exposures,  
typically of  300, 45 and  25\,s, complemented by a number of  
short exposures (typically a few second long) in order to  
properly measure the brightest stars.  
  
Similar observations (i.e similar strategy and spatial coverage)    
were obtained on June 2002 with  
the {\it Advanced Camera for Survey} (ACS) on board 
HST through three filters [namely F625W ($R_{625}$), F435W
($B_{435}$), F658N  (H$\alpha$)] within a programme devoted to the 
search for  Cataclysmic Variables in $\omega$~Cen. 
We retrieved three 340\,s  exposures in the $B_{435}$ and $R_{625}$ 
filters for each of the nine pointings
from the {\it  ESO/ST-ECF Science  Archive}. 
Two additional short exposures of 8 and 12\,s, in the  
$R_{625}$ and $B_{435}$ bands, respectively, were also retrieved in order to  
properly measure the brightest stars.   
  
Both the FORS1 and ACS datasets have been reduced with DAOPHOT~II (Stetson 1987,  
1994) using the same procedure. To optimize the search for faint  
objects, the source detection has been performed on the combined  
median image, obtained from the deepest exposures of each  
pointing. The list of detected objects in the combined
image has then been fed to the PSF-fitting routines of DAOPHOT 
and ALLSTAR that were run on each exposure, separately. 
Each ACS pointing was corrected for geometric  
distortions using the prescriptions by  Hack \& Cox (2001).  
The resulting  
instrumental magnitudes have been converted into a common  
(instrumental) photometric system and averaged together.  
The final FORS1 and ACS instrumental catalogs have then  
been photometrically calibrated: the former, by using 
a large sample of stars in common with the WFI catalog
by Pancino et al. (2000) (see also Sollima et al. 2004 for more details), 
while the latter has been transformed 
into  the VEGA-MAG system by using preliminary zero-points, 
kindly  provided by G. De Marchi (private communication).
 
   
\section{Results}  
\label{sec:res}  
  
\begin{figure}   
\plotone{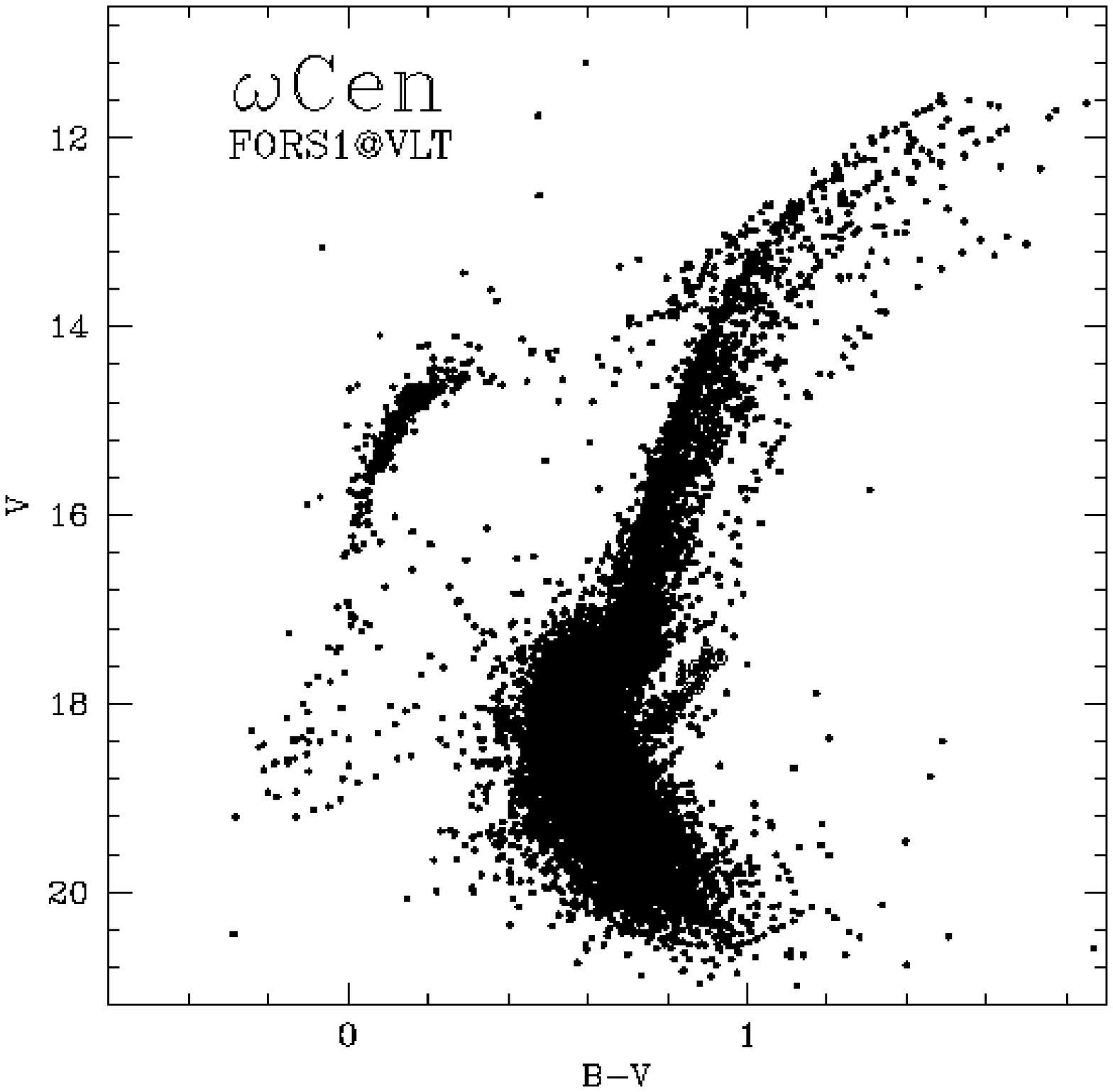}   
\caption{ $(V,B-V)$ CMD for stars identified in the nine
fields observed with FORS1 at the ESO/VLT.
About 14,000 stars with the lowest photometric error
have been plotted. 
The SGB-a population, merging into the cluster MS, is clearly
visible. Stars marked with large open circles are
identified as SGB-a stars in both the FORS1 and ACS
catalogs. \label{fig:CMD1}}  
\end{figure}

Figure \ref{fig:CMD1} shows the $(V,B-V)$ 
CMD obtained from the ground-based FORS1 VLT  
observations. 
Photometric errors have  been calculated for each star from the r.m.s.
scatter of repeated exposures. In order to minimize the probability
of photometric contamination by nearby stars, only the 14,000 stars with
the lowest photometric errors ($\sigma_V\simeq\sigma_B<$0.01 mag,
see Sollima et al. 2004, in preparation) have been plotted.
The complex structure of the RGB shows beautifully, but the most  
striking feature in this CMD is the additional, narrow SGB  
sequence, which sticks out of the main SGB-TO region of the  
cluster. 

\begin{figure}   
\plotone{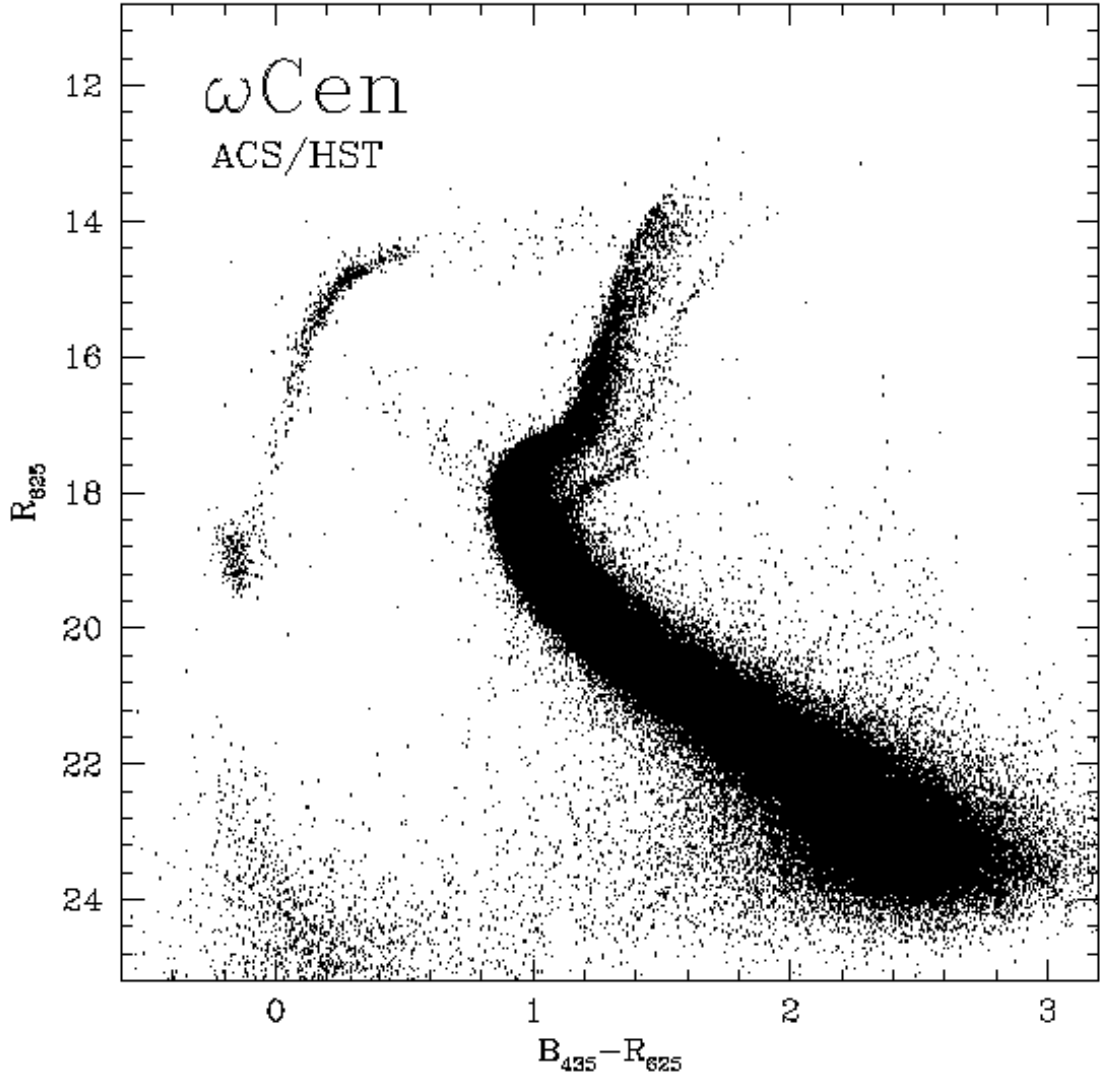}   
\caption{$(R_{625},B_{435}- R_{625})$ CMD for more than 
400,000 stars identified in the nine ACS fields. The SGB-a is 
clearly visible, along with the complex sub-structures 
of the TO-SGB region of $\omega$~Cen. \label{fig:CMD2}}  
\end{figure}

Fig.~\ref{fig:CMD2} displays the $(R_{625},B_{435}-R_{625})$ CMD   
obtained from the HST/ACS sample: more than 400\,000  
stars with the lowest photometric errors are plotted. 
 All the evolutionary sequences are  
extremely well defined, apart from the brightest giants that are  
saturated even in the shortest exposures (not plotted in 
Fig.~\ref{fig:CMD2}).
The SGB-a is better defined than in the 
FORS1 CMD, and the complex SGB/TO  
structures are more clearly distinguishable. 
In this CMD, the SGB-a appears as the direct  
continuation of the RGB-a, although 
the presence of a number of stars located in between the 
main SGB and the SGB-a (at $R_{625}=17-18$) offers room to
other possible interpretations. 
The SGB-a sequence merges  
into the bulk of the cluster MS at a significantly fainter  
magnitude ($\Delta V \sim \Delta R_{625} \sim 0.6$\,mag) 
with respect to the dominant  
population.    
SGB-a stars in common to both the FORS1 and ACS catalogs 
are marked in Figure 1.
  
Both the FORS1 and ACS CMDs 
show that the MS-TO of the SGB-a population  
($R_{625}^{TO}(SGB-a)=18.40 \pm0.15$) is significantly fainter  
than the MS-TO of the dominant cluster population  
($R_{625}^{TO}(MP)=17.80 \pm0.15$).   A hint of the
presence of such a feature was already noted
by Anderson (2002, see his Fig.~1), in the core of the
cluster on the basis of an instrumental 
$(U_{336},U_{336}-R_{675})$ CMD obtained from WFPC2/HST observations. 
However,
the high quality CMDs presented here provide the first clear-cut
definition of this sequence in the CMD of  $\omega$~Cen.

The TO and SGB regions of $\omega$~Cen show the same level of
complexity already observed at the RGB level: 
{\it (i)}~a well defined and populated main SGB sequence,  
highly homogeneous both in terms of age and  metallicity. 
{\it (ii)})~a more continuous distribution of stars toward the red,  
tracing a less homogeneous stellar component, with different 
metallicities and/or ages.
{\it (iii)}~a well defined narrow additional SGB, 
redder and fainter than the 
main SGB. The sharpness of this sequence  
strongly suggests a high degree of 
homogeneity in the age and metallicity of its stars.

\section{Discussion}  
\label{sec:dis}  
  
In the simplest scenario, the SGB-a is the extension of the RGB-a.
Following this hypothesis, we assume 
that the anomalous branches are 
constituted by the same population of metal-rich stars.
  
We then performed a detailed comparison with theoretical models 
by using a set of theoretical 
isochrones calculated adopting the most 
up-to-date input  physics (Straniero, Chieffi \& Limongi 1998). 
In particular, the equation of state includes the electrostatic 
correction (see Prada Moroni \& Straniero 2002, for a detailed 
description) and microscopic diffusion (gravitational settling 
and thermal diffusion). 
The theoretical isochrones have been 
transformed into the observational planes by means of the 
synthesis code described in Origlia \& Leitherer (2000), using 
the model atmospheres by Bessell, Castelli \& Plez (1998). 
For the ACS data set the 
filter responses and camera throughputs, kindly provided by 
the ACS User Support Team, have been used.   
 
In order to perform a meaningful comparison between theoretical  
isochrones and observed sequences, we need to adopt an average  
metallicity for each population. As discussed in previous  
papers (see Ferraro et al. 1999, 2000 and references therein) an  
accurate comparison requires the use of the {\em global  
metallicity}, [M/H], obtained by combining the classical [Fe/H]  
abundance with the contribution of the $\alpha-$elements.  
Detailed abundances for the different stellar  
populations in $\omega$~Cen have been published by several authors  
(see, e.g., Norris \& Da Costa 1995; Smith et al. 2000; Vanture  
et al. 2002), including the recent studies on the RGB-a  
stars published by our group (Pancino et al. 2002, 2003; Origlia et
al. 2003). These papers show that,   
while the metal poor populations at [Fe/H]$\le-1.0$ 
turn out to be $\alpha$-enhanced
([$\alpha$/Fe]$\simeq0.3-0.4$~dex) 
compared to the solar values, the most metal rich one is only marginally 
$\alpha$-enhanced [$\alpha$/Fe]$\le0.1-0.2$~dex.  

Using the above values and the prescription by Salaris, Chieffi  
\& Straniero (1993, see their eq.~3), we adopted a global metallicity  
[M/H]$\simeq-1.5$ (or $Z=0.0006$) for the metal-poor, dominant  
population and [M/H]$\simeq-0.6$ (or $Z=0.005$) for the  
metal-rich population.  
The  mean interstellar extinction coefficients listed in
Table 2 by Savage \& Mathis (1979) have been adopted.
  
By best-fitting the $Z=0.0006$ isochrone to
the metal-poor component, we obtained 
$(m-M)_0=13.70$ and $E(B-V)=0.11$ and an age of
$15$\,Gyr. These values  are in good agreement with the most
recent determinations by Thompson et al. (2001)
and Lub (2002) ($(m-M)_0=13.65 \pm 0.11$ 
and $E(B-V)=0.11\pm0.02$) and the most recent age determination by 
Gratton at al. (2003).
By using these values, the position of
the anomalous SGB has been compared with a set of $Z=0.005$ 
isochrones in the 12-18 Gyr age range. 
It is worth of mentioning that, since we are performing a
strictly differential analysis between the SGB-a and the
dominant cluster population, small uncertainties in the
reddening and distance modulus would not affect the overall results.
 
\begin{figure}   
\plotone{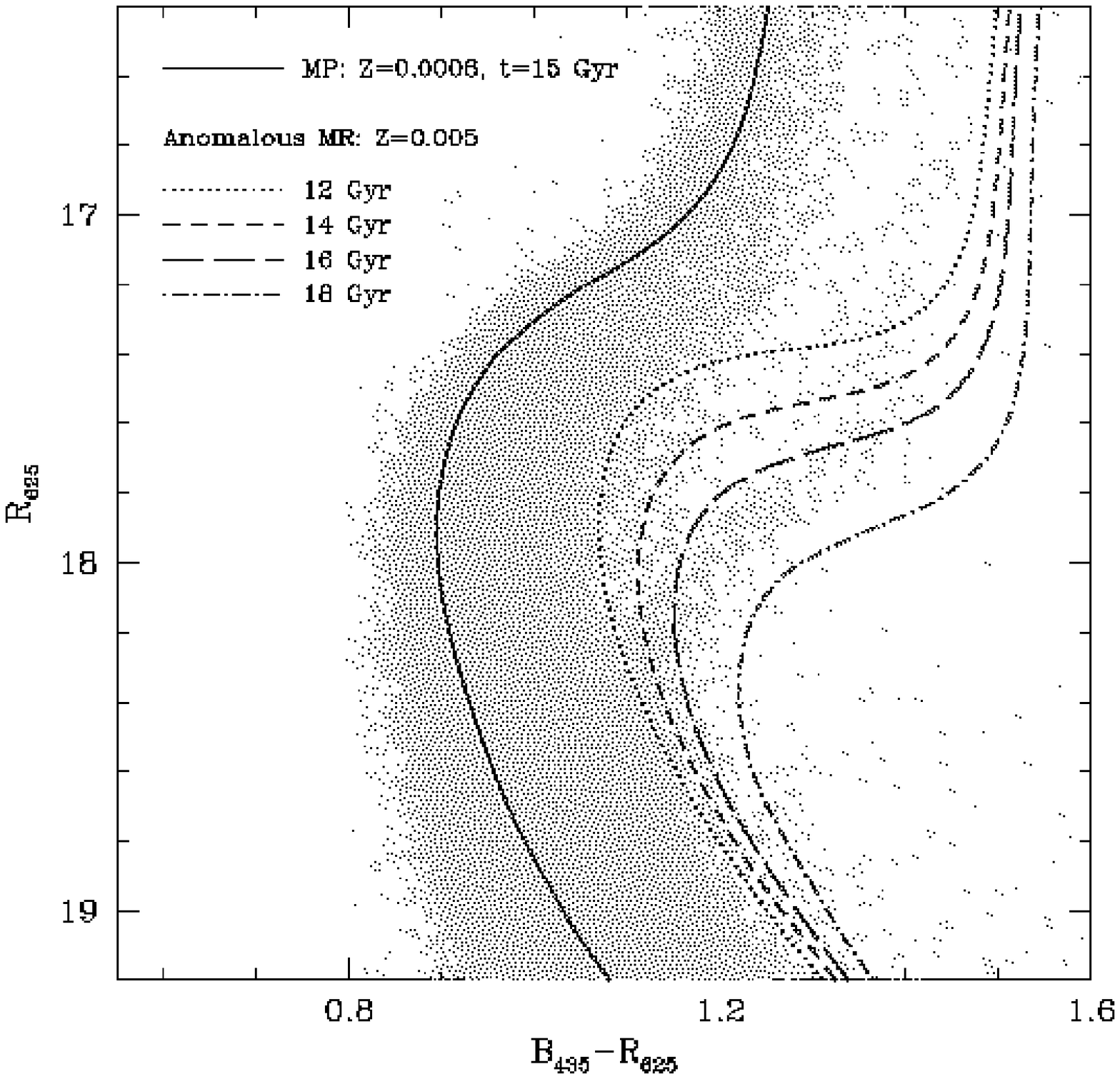}   
\caption{ 
Zoomed $(R_{625},B_{435}- R_{625})$ CMD of the TO region. 
Two isochrone sets are overplotted. 
The metal-poor dominant population is well 
reproduced by a $Z=0.0006$ isochrone of 15 Gyr (heavy solid 
line). A set of 
$Z=0.005$ isochrones in the 12-18 Gyr age range  
is also shown.
\label{fig:fit1}}  
\end{figure}

As can be appreciated in Fig.~\ref{fig:fit1}, while the metal-poor  
population is reasonably well reproduced, 
the TO level of the anomalous  population is 
definitely fainter than what predicted by a 
metal-rich isochrone significantly younger ($t<14 Gyr$)
than the dominant population. 

A detailed analysis of Fig.~\ref{fig:fit1} shows that indeed the
observed morphology of the anomalous SGB
 is not correctly reproduced by any
$Z=0.005$ isochrone, regardless of age. 
In fact, though the MS-TO level of the anomalous population appears consistent
with an age of 17 Gyr, (i.e. $\delta t \sim 2$  Gyr older than the 
metal-poor component), the TO color is
significantly bluer than the corresponding isochrone.
Moreover, the morphology, extension and position of the
lower RGB-a and SGB-a are not correctly reproduced, since 
{\it 1)}~the SGB-a appears significantly less extended in color and much
steeper than what predicted by a metal rich isochrone
(somewhat suggestive of a lower metal content);  
{\it 2)}~the position of the base of the RGB-a 
appears significantly bluer than what predicted
by the isochrone.\footnote{However, when evaluating the
importance of this discrepancy one needs to keep in
mind that the position of the RGB in the theoretical models
is highly uncertain due to the mixing-length convection
approximation.}
The same problems have been noted by fitting the SGB-a in
the (V,B-V) plane.
Note that the same isochrone set nicely fits the
SGB/TO region of well-known clusters of similar
metallicity (as 47 Tuc).

Other two parameters,  
namely the He and CNO abundances can 
affect the location and shape of the SGB-TO region in the CMD.
In order to evaluate the effect of an
enhanced He abundance over the shape of the SGB, we computed
a set of suitable models at Y=0.28 ($\delta
Y=0.05$ with respect to the standard value adopted above).
From this models we noted that
 an increase of the He abundance does not
significantly affect the TO level while it decreases 
the SGB extension by moving blueward the location of the RGB by 
few hundredths of magnitude
in the $B-R$ color. Hence an increase of the He abundance 
cannot be invoked to explain the faintness of the TO of
the anomalous population.
 
Renzini (1977, see also the discussion in Salaris, Chieffi \&
Straniero 1993) showed that the TO luminosity and temperature 
primarily depend on the CNO and Ne abundances, while the RGB colors 
mainly depend on the Fe, Si and Mg ones. 
Thus an {\it ad hoc} increase of $Z_{C,N,O,Ne}$ would
{\it (i)}~move the
TO color towards the red while leaving  the
RGB location unchanged, thus reducing the SGB extension and
 {\it (ii)}~move the TO towards fainter magnitudes
(see Fig.~22 of Salaris, Chieffi \& Straniero 1993). 
This goes  exactly in the direction of moving the young ($\approx$12 Gyr) 
isochrone towards the observed sequence. 
However, Salaris, Chieffi \& Straniero (1993) 
quantified the effect of O 
(the most abundant element among the CNO group and Ne) 
overabundance over the TO luminosity:
an O overabundance by a factor of 4 would increase the
TO level by $\delta V \sim 0.07$ mag. 
An unlikely O overabundance of $\approx$ 20 would be necessary
to move the 12 Gyr isochrone to the level of the observed TO. 
Note that spectroscopic [O/Fe] measurements of a few RGB-a stars 
(see Fig.~5 in Origlia et al. 2003) indicate only
marginal (if any) O overabundance.    

In summary, it seems impossible to reconcile the TO level of a  
young ($\approx 12$ Gyr) metal rich isochrone with the observed 
feature.
Accurate measurements of C,N,Ne and O abundances for a 
significant number of SGB-a stars are urgently needed to check for 
possible anomalies, which could play some role in characterizing the 
morphology and the position of the SGB-a.
Moreover, additional modeling with a detailed fine-tuning among  
[M/H], CNO and Ne abundances could indeed
improve the agreement between old (14-16 Gyr) isochrones and the 
observed SGB-a morphology.

A number of papers (Lee et al. 1999, Hughes \&
Wallerstein 2000, Hilker \& Richtler 2000, Rey et al. 2003) argue, from
indirect evidences, that the age of the most metal rich stellar population in
$\omega$Cen is significantly ($\Delta t \sim 4 $ Gyr) younger that the most
metal poor one. 
Under the hypothesis that the SGB-a is related to the RGB-a, 
we demonstrated from the direct detection of the TO, 
that this anomalous, metal rich population 
cannot be younger than the most metal poor one. 
This evidence excludes the possibility that such a 
metal rich population be the latest product of  
the self-enrichment process in $\omega$ Cen (as suggested
by  Rey et al. 2003 and references therein), while  
it would support its extra-cluster origin 
(see Ferraro, Bellazzini, Pancino 2002).  

In the scenario proposed by Bekki
\& Freeman (2003), $\omega$~Cen is the complex
stellar relic of a nucleated dwarf galaxy that merged in a
remote epoch with the Galaxy (see also Dinescu et al.
1999;  Majewski et al. 2000). Populations with different
metallicities would be the result of subsequent radial
gas inflows into the main body of the parent dwarf galaxy.
This scenario can reasonably account for  
the genuine $\omega$~Cen sub-populations at [Fe/H]$\lesssim-1.0$.
Conversely, the anomalous, metal rich population could be a small 
stellar system 
(a globular cluster?) accreted by the main body of $\omega$~Cen 
during the disrupting interaction with the Galaxy.
The assumption that $\omega$~Cen could host a satellite
globular cluster
is not fully {\em ad hoc}, since both the other two known
disrupting systems in the Galaxy (the Sagittarius dwarf
spheroidal -- Ibata et al. 1994 and the Canis Major dwarf
galaxy -- Ibata et al. 2003) have their own globular cluster
system.

A number of alternative hypothesis about the nature of the 
SGB-a become possible if this population is not related  to
the RGB-a.  The SGB-a observed morphology
could be indeed reproduced  by old, low-metallicity isochrones, 
but this would require the use of a different
distance  and reddening with respect to main cluster
population. Such an hypothesis would imply that the
anomalous SGB trace a metal-poor stellar system, in the
background of the main body of $\omega$~Cen.  Presently we
have no possibility to discriminate between the various 
scenarios. Only direct measurements of radial velocities
and metal abundances of a significant sample of SGB-a
stars  will allow us to answer the enigma posed by the
discovery of this new feature.

\acknowledgements  
  
We warmly thank Guido De Marchi for providing the photometric 
zero-points  for the ACS filters and Bob Rood and Flavio Fusi
Pecci for helpful comments and suggestions. 
We also thank  the Referee, Russell Cannon,   for his 
comments and suggestions.
This research was supported by the Agenzia Spaziale Italiana (ASI) and the 
Ministero dell'Istruzione, dell'Universist\`a e della Ricerca. 
This research has made use of 
the ESO/ST-ECF Science 
Archive facility which is a joint collaboration of the European 
Southern Observatory and the Space Telescope - European 
Coordinating Facility.

\clearpage

\clearpage  
   
\clearpage  
   
\begin{thebibliography}{}  

\bibitem[Anderson (2002)]{A02} Anderson J., in 
       "Omega Centauri: A Unique Window into Astrophysics", 2002, 
       ed. F. van Leeuwen, J.D.Huges \&  G. Piotto, ASP Conf.Series, 87
\bibitem[Anderson (2003)]{A03} Anderson J., 
in "New Horizon in globular cluster Astronomy", 2003, ed. G.Piotto,
         G. Meylan, S.G. Djorgovski \& M. Riello, ASP Conf.Series, 125  
\bibitem[Bekki \& Freeman (2003)]{}  Bekki, K., Freeman, K.C.,2003,   
         astro-ph/0310348  
\bibitem[Bessel, Castelli \& Plez  (1998)]{}  
      Bessel, M. S., Castelli, F., \& Plez, B. 1998, \aap, 333, 231    
\bibitem[Binney, Gherard \& Silk (2001)]{bgs01} Binney, J., Gherard, O. \& Silk,
         J., 2001, MNRAS, 321, 471    
\bibitem[Dinescu, Girard \& van Altena(1999)]{dinescu}  Dinescu, D.I.,   
         Girard, T.M. \& van Altena, W.F.\ 1999, \aj, 117, 1792     
\bibitem[Ferraro et al. (1999)]{F99} Ferraro, F.R., Messineo, M., Fusi Pecci, F., De Palo, M.A., Straniero, O.,
         Chieffi, A. \& Limongi, M., 1999, AJ, 118, 1738
\bibitem[Ferraro et al. (2000)]{F00} Ferraro F.R., Montegriffo P.,
         Origlia L.,Fusi Pecci F., 2000, AJ 119, 1282-1295	 
\bibitem[Ferraro, Bellazzini \& Pancino (2002)]{moti}  
         Ferraro, F.R., Bellazzini, M. \& Pancino, E., 2002, ApJ, 573, L95  	   
\bibitem[Gratton et al. (2003)]{G03} Gratton, R.G., Bragaglia,
A., Carretta, E., Clementini, G.,
 Desidera, S., Grundahl, F., Lucatello, S., 2003, \aap, 408, 529 
\bibitem[Hack \& Cox (2001)]{hk01} Hack, W. \& Cox,  
         C., 2001, Instrument Science Report 2001-008  
\bibitem[Hilker \& Richtler(2000)]{hk00} Hilker, M. \& Richtler,  
         T., 2000, \aap, 362, 895  
\bibitem[Hughes \& Wallerstein(2000)]{hughes} Hughes, J. \&  
         Wallerstein, G., 2000, \aj, 119, 1225  
\bibitem[Ibata et al. (1994)]{igi94} Ibata R. A., Gilmore, G., Irwin, M. J.,
         1994, Nature, 370, 194
\bibitem[Ibata et al. (2003)]{I03} Ibata R.A., Irwin, M.J., Lewis, G.F.,
         Ferguson,  A.M.N., Tanvir, N., 2003, MNRAS, 340, 21	 
\bibitem[Lee et al.(1999)]{lee} Lee, Y.-W.,Joo, J.-M.,Sohn, Y.-J.,      
         Rey, S.-C., Lee, H.-C. \& Walker A. R.       
         1999, \nat, 402, 55 
\bibitem[Lub (2002)]{l02} Lub, J., 2002, in $\omega$
          Centauri: a unique window into Astrophysics, F. van
          Leeuwen, J. Hughes and G. Piotto, eds., S. Francisco, ASP
          Conf. Ser., 265, 95
\bibitem[Meylan et al.(1995)]{M95} Meylan, G., Mayor, M., 
         Duquennoy, A., \& Dubath, P. 1995, A\&A, 303, 761
\bibitem[Majewski et al.(2000)]{maj} Majewski, S.R., Patterson, R.J., Dinescu,  
         D.I., Johnson, W.Y., Ostheimer, J.C., Kunkel, W.E., Palma, C., 2000,   
         in The Galactic Halo: From Globular Cluster to Field Stars, N.A.   
         Magain, D. Caro, G. Parmentier, and A.A. Thul eds., (Li\`ege,  
         Belgique: Institut d'Astrophysique et de Geophysique), p. 619            
\bibitem[Norris \& Da Costa(1995)]{norris} Norris, J.E., Da Costa,  
         G.S. 1995, \apj, 447, 680   
\bibitem[Norris et al.(1997)]{nor97} Norris, J.E., Freeman, K.C., Mayor, M.   
        \& Seitzer, P., 1997, \apj, 487, L187 (N97)      
\bibitem[Origlia \& Leitherer (2000)]{}  
         Origlia, L., \& Leitherer, C. 2000, \aj, 119, 2018  
\bibitem[Origlia et al.(2003)]{} Origlia, L., Ferraro, F.R.,  
          Pancino, E., Bellazzini, M., 2003, ApJ, 591, 916  
\bibitem[Pancino et al.(2000)]{p00} Pancino, E., Ferraro, F.R., Bellazzini,   
         M., Piotto G., \& Zoccali, M., 2000, \apjl, 534, L83-L87 (P00)      
\bibitem[Pancino et al.(2002)]{p02} Pancino, E., Pasquini, L., Hill, V.,   
         Ferraro, F.R. \& Bellazzini, M., 2002, \apjl,   
	 568, L101      
\bibitem[Pancino et al.(2003)]{russo} Pancino, E., Seleznev, A.,   
         Ferraro, F.R., Bellazzini, M. \& Piotto, G. 2003,
         MNRAS, 345, 683
\bibitem[Pancino (2003)]{} Pancino, E., 2003, PhD. Thesis,
         Bologna University.
\bibitem[Prada Moroni \& Straniero (2002)]{pms02} Prada Moroni, P. \& Straniero,
         O., 2002, ApJ, 581, 585 
\bibitem[Thompson et al.(2001)]{ogle} Thompson, I., Kaluzny, J., Pych, W.,  
         Burley, G., Krzeminski, W., Paczynski, B., Perrson, S.E., Preston,   
	 G.W., 2001, \aj, 121, 3089        
\bibitem[Rey et al.(2003)]{rey03} Rey S.-C., Lee Y.-W., Ree C. H., Joo
         J.-M., Sohn Y.-J. \& Walker, A. 2003, AJ in press (astro-ph/0310773)
\bibitem[Renzini (1977)]{}Renzini, A., 1977, in Advanced
         Stages in Stellar Evolution, ed. P. Bouvier \& A.  Maeder
         (Geneva: Geneva Obs.), 151
\bibitem[Salaris, Chieffi \& Straniero (1993)]{SCS93} Salaris, M., Chieffi, A., Straniero, O., 1993, 
          \apj, 414, 580
\bibitem[Savage \& Mathis (1979)]{SM79} Savage, B.D., Mathis, J.S., 1979, ARA\&A, 17, 73S
\bibitem[Smith et al.(2000)]{smith} Smith, V.V., Suntzeff, N.B.,  
         Cunha, K., Gallino, R., Busso, M., Lambert, D.L., Straniero, O.,   
	 2000, \aj, 119, 1239
\bibitem[Stetson (1987)]{S94} Stetson, P. B., 1987, PASP, 99, 191
\bibitem[Stetson (1994)]{S94} Stetson, P. B., 1994, PASP, 106, 250
\bibitem[Straniero, Chieffi \& Limongi (1997)]{scl97} Straniero O.,
        Chieffi A. \& Limongi M., 1997, ApJ, 490, 425
\bibitem[Suntzeff \& Kraft(1996)]{sk96} Suntzeff, N.B.,  \& Kraft R.P., 1996,   
         \aj, 111, 1913
\bibitem[Thompson, et al (2001)]{} Thompson, I.B., Kaluzny, J., 
         Pynch, W., Burley, G., Krzeminski, W., Paczynski, B.,
         Persson, S.E. \& Preston, G.W., 2001, AJ, 121, 3089             
\bibitem[van Leeuwen et al.(2000)]{vanl00} van Leeuwen, F., Le Poole, R.S.,   
         Reijns, R., Freeman, K.C. \& De Zeeuw, P.T., 2000, A\&A, 360, 472   
\bibitem[Vanture et al (2002)]{} Vanture, A.D., Wallerstein, G., 
         \& Suntzeff, N.B., 2002, ApJ, 569, 984     
\end{thebibliography}
\end{document}